\documentclass[reprint,amsmath,amssymb,aps,prl]{revtex4-2}

\usepackage{graphicx}% Include figure files
\usepackage{dcolumn}% Align table columns on the decimal point
\usepackage{bm}% bold math
\usepackage{epstopdf}
\usepackage[mathlines]{lineno}% Enable numbering of text and display math
\usepackage{harpoon}
\usepackage{makecell}
\usepackage{subfigure}
\usepackage{hyperref}
\usepackage{color}
\usepackage{braket}
\usepackage{amsmath}
\usepackage{mathrsfs}
\usepackage{appendix}
\usepackage{multirow}
\usepackage{float}
\usepackage{enumerate}
\usepackage{ulem}
\usepackage{slashed}

\begin{document}

\title{New Limits on Ultralight Axionlike Dark Matter from Reanalyzed Data}

\author{K. Y. Zhang}
\affiliation{Institute of Nuclear Physics and Chemistry, China Academy of Engineering Physics, Mianyang, Sichuan 621900, China}

\author{L. Y. Wu}
\affiliation{Institute of Nuclear Physics and Chemistry, China Academy of Engineering Physics, Mianyang, Sichuan 621900, China}
\affiliation{Key Laboratory of Nuclear Physics and Ion-beam Application (MOE), Fudan University, Shanghai 200433, China}

\author{H. Yan} \email{h.y\_yan@qq.com}
\affiliation{Institute of Fundamental Physics and Quantum Technology, and School of Physical Science and Technology, Ningbo University, Ningbo, Zhejiang 315211, China}

\begin{abstract}
 New limits on the axion-nucleon coupling over the axion mass region $10^{-24} \leq m_a \leq 5 \times 10^{-21}$ eV are derived by reanalyzing data from laboratory measurements on Lorentz and $CPT$ violation. These results establish the first laboratory constraints on the axion-nucleon coupling for axion masses below $10^{-22}$ eV. For $10^{-22} \leq m_a \leq 5 \times 10^{-21}$ eV, the results improve upon previous laboratory limits by more than 3 orders of magnitude, exceeding for the first time the astrophysical limits from supernova SN1987A cooling. For the axion mass range of interest corresponding to ultralow frequencies, the crucial local phase of the axion field is considered. Furthermore, the obtained limits are nearly equivalent to those projected for a recently proposed experiment employing high-intensity neutron beams at the European Spallation Source. For an alternative type of axion-nucleon interaction, the quadratic wind coupling, the constraints exceed the current best results by approximately 2 orders of magnitude.
\end{abstract}

\date{\today}

\maketitle

\textit{Introduction.}---According to astrophysical observations, dark matter accounts for approximately 27\% and dark energy about 68\% of the universe's total energy density, leaving only 5\% composed of ordinary matter. However, the direct detection of dark matter—typically based on its nongravitational interactions with particles and fields of the Standard Model—has remained an ambitious goal for several decades \cite{Liu2017NatPhys, Bertone2018Nature, Bertone2018RMP}. Successfully achieving this goal would not only unveil the nature of dark matter but also provide crucial insights into new physics beyond the Standard Model \cite{Safronova2018RMP}. One of the most well-motivated candidates for dark matter is the axion \cite{Bertone2005PhysRep, Yannis2022SciAdv}, which was initially proposed to resolve the strong $CP$ problem of quantum chromodynamics (QCD) \cite{PQ1977PRL, PQ1977PRD, Wilczek1978PRL, Weinberg1978PRL}. Beyond the QCD axion, a variety of axionlike particles (hereafter generically referred to as axions) have been predicted in generalized theories where the PQ (Peccei-Quinn) symmetry is broken at different energy scales \cite{Svrcek2006JHEP, Arvanitaki2010PRD}.

Laboratory searches for axion dark matter have focused on the possible axion-photon, axion-gluon, and axion-fermion couplings. For instance, photons can be created by the conversion of axions in strong electromagnetic fields via the Primakoff effect \cite{Sikivie1983PRL, Bradley2003RMP, Asztalos2010PRL, Brubaker2017PRL}. Interactions of the coherently oscillating axion dark matter field with gluons and fermions can induce oscillating electric dipole moments (EDMs) of nucleons and atoms \cite{Stadnik2014PRD, Roberts2014PRL}.
Nuclear magnetic resonance (NMR) techniques have been employed to search for anomalous dark-matter-driven spin precession \cite{Wu2019PRL, Garcon2019SciAdv, Aybas2021PRL}.
In addition, axions might mediate exotic spin-dependent interactions between macroscopic objects through the axion-fermion coupling \cite{Moody1984PRD, Sikivie2021RMP, Wu2022PRL, Wu2023PRL, Wu2024JHEP}.

Since the PQ symmetry can be broken at arbitrarily large energy scales, the axion mass is theoretically unconstrained \cite{Svrcek2006JHEP}. A lower mass bound, $m_a \gtrsim 10^{-22}$ eV, arises from the requirement that, if axions saturate the observed cold dark matter content, their de Broglie wavelength must not exceed the size of the dark matter halos of the smallest dwarf galaxies \cite{Bozek2015, Schive_2016}. However, axions with masses below this bound can still constitute a partial fraction of dark matter \cite{Marsh2016PhysRep}. Several experimental proposals and searches have targeted this regime of extremely ultralight dark matter \cite{Van2015PRL, Arvanitaki2015PRD, Hees2016PRL}. Notably, axions with ultralow masses in the range $10^{-24} \lesssim m_a \lesssim 10^{-20}$ eV have been proposed as dark matter candidates that are observationally distinct from and, in some scenarios, potentially favorable compared to the archetypal cold dark matter model  \cite{Hu2000PRL, Marsh2013MNRAS, Schive2014NatPhys}.

Recently, constraints on the coupling of axion dark matter to gluons in the axion mass range $10^{-24} \leq m_a \leq 10^{-17}$ eV have been set by the PSI neutron EDM experiment \cite{Abel2017PRX}. For $10^{-23} \lesssim m_a \lesssim 10^{-18}$ eV, limits on the axion-electron coupling were derived from experimental data analyzed using a rotating torsion-pendulum \cite{Terrano2019PRL}. In the range $2 \times 10^{-23} \leq m_a \leq 4 \times 10^{-17}$ eV, the axion-antiproton interaction has been constrained through analysis of the antiproton spin-flip resonance \cite{Smorra2019Nature}.

Very recently, Ref.~\cite{Fierlinger2024PRL} proposed a neutron beam experiment at the European Spallation Source (ESS) using Ramsey interference techniques to search for ultralight axion dark matter. This experiment leverages the coupling between axions and neutron spins, utilizing the high-intensity neutron beam from the ESS HIBEAM line and the Ramsey-separated oscillating field method. By comparing the neutron spin precession frequency to an external reference frequency, the experiment aims to detect frequency shifts induced by axion dark matter. For a one-year runtime, the sensitivity is expected to improve by 2–3 orders of magnitude, covering the axion mass range $10^{-22}$ eV to $10^{-16}$ eV.

In this Letter, the first constraints on the axion-nucleon coupling for axion masses below $10^{-22}$ eV are presented, and for the mass range $10^{-22} \leq m_a \leq 5\times 10^{-21}$ eV, new limits comparable to the sensitivity expected from the aforementioned ESS experiment \cite{Fierlinger2024PRL} are established.
The results also surpass previous laboratory limits by more than 3 orders of magnitude and represent the first laboratory constraints that exceed the astrophysical bounds derived from supernova SN1987A cooling.

\textit{The basic idea.}---An axion field might interact with nucleons via the derivative coupling:
\begin{equation}
\mathcal{L}_{\mathrm{int}} = g_{\mathrm{aNN}} \partial_\mu a \bar N \gamma^\mu \gamma^5 N,
\end{equation}
where $N$ represents the nucleon field and $\bar N$ its Dirac adjoint, and $g_{\mathrm{aNN}}$ is the coupling strength.
Axions that could have been produced in the early universe manifest as a classical field \cite{Preskill1983PLB, Abbott1983PLB, Dine1983PLB}
\begin{equation}
a(t) = a_0 \cos (2\pi\nu_a t + \phi),
\end{equation}
oscillating at the axion's Compton frequency $\nu_a = m_a c^2/h$, where $c$ is the speed of light and $h$ is the Planck constant.
The field amplitude $a_0$ can be estimated by the Galactic dark-matter energy density $\rho_a = m_a^2a_0^2c^2/(2\hbar^2)\approx 0.4$ GeV cm$^{-3}$ \cite{Catena2010JCAP}.
The phase of the field $\phi$ can be a random number ranging from 0 to $2\pi$.
Then, the spatial components of the derivative coupling of the axion field with nuclear spins simplify in the nonrelativistic limit as follows:
\begin{equation}
H_{\mathrm{int}} = 2g_{\mathrm{aNN}} \sqrt{2\hbar^3 c\rho_a} \sin(2\pi\nu_a t +\phi) \bm{v}_a \cdot \bm{I}_N,
\label{Hint}
\end{equation}
where $\bm{v}_a$ represents the expected average axion wind velocity, and $\bm{I}_N$ denotes the nuclear spin.
This interaction is in analogy to the Zeeman interaction of $\hbar\gamma \bm{B}_a \cdot \bm{I}_N$, where an effective magnetic field induced by axion dark matter reads
\begin{equation}
\bm{B}_{a} = \frac{2g_{\mathrm{aNN}}}{\gamma} \sqrt{2\hbar c\rho_a} \sin(2\pi\nu_a t +\phi) \bm{v}_a,
\label{beff}
\end{equation}
and $\gamma$ is the gyromagnetic ratio of the nuclear spin.
Consequently, by detecting the effects of this field on nuclear spin, the strength $g_{\mathrm{aNN}}$ can be measured or constrained.
We note that the phase $\phi$ has occasionally been omitted in previous literature.
At the same time, it could be critically important when the axion field frequency under study is extremely low--for instance, as low as $10^{-8}$ Hz for $m_a\sim 10^{-22}$ eV.
In Ref. \cite{Wu2019PRL}, $|\sin\phi|$ is approximated by its average value in the analysis for field frequencies below $3\times10^{-7}$ Hz.
In Refs. \cite{Abel2017PRX,Terrano2019PRL}, a linear combination of sine and cosine functions with free-fit amplitude parameters is considered, which is equivalent to a single sinusoidal function of unknown amplitude and offset phase, like in Eq. \eqref{beff}.

\begin{figure}[htbp]
    \centering
    \includegraphics[width=0.5\textwidth]{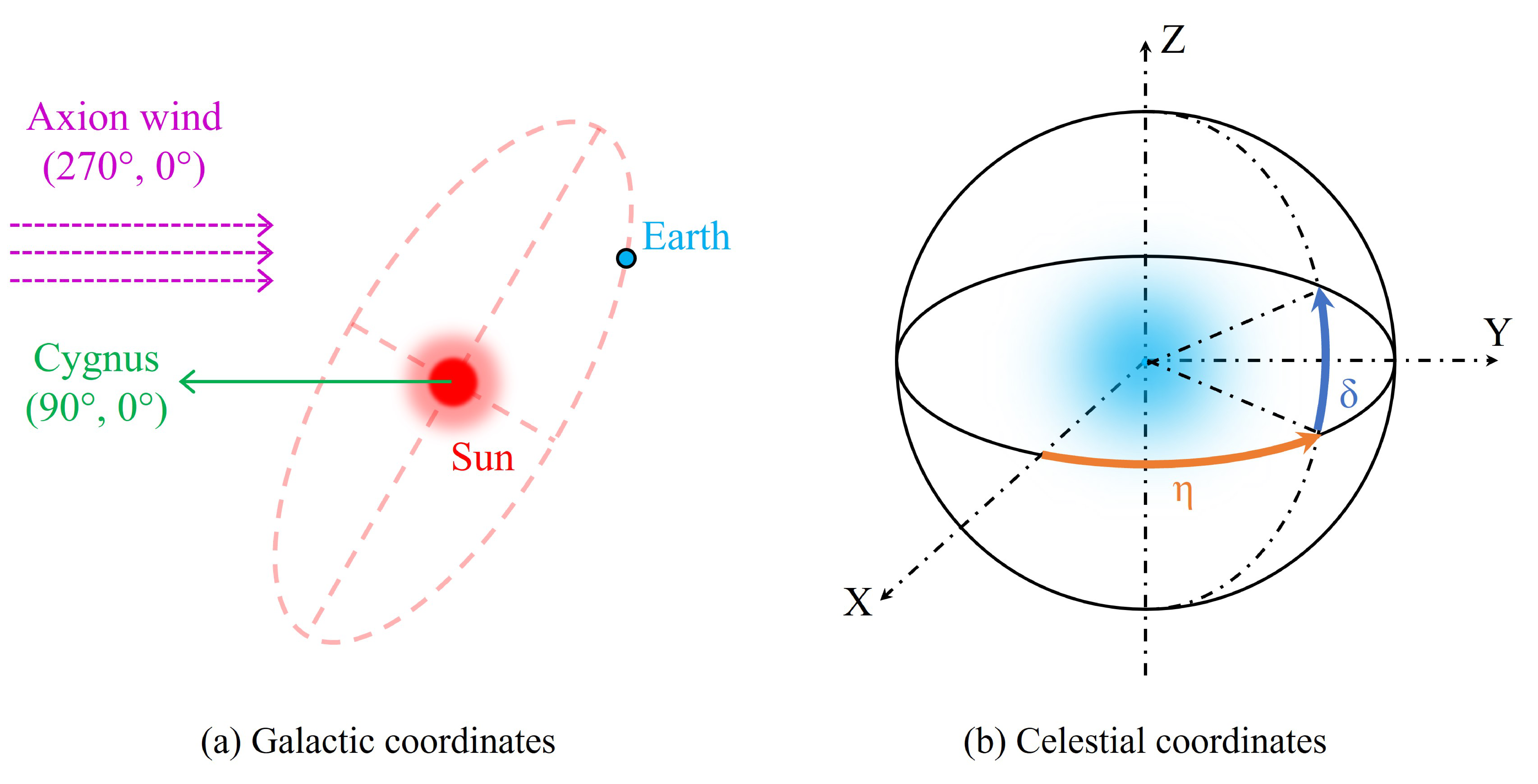}
    \caption{Galactic coordinates (a) and celestial coordinates (b). In (a), the Sun moves toward the Cygnus constellation with $90^\circ$ longitude and $0^\circ$ latitude, and the velocity direction of the axion wind is $270^\circ$ longitude and $0^\circ$ latitude. When transforming to (b), the velocity direction can be defined by declination $\delta$ and right ascension $\eta$.}
    \label{fig1}
\end{figure}

In Galactic coordinates, as shown in Fig. \ref{fig1}(a), the Sun is moving towards the Cygnus constellation with $90^\circ$ longitude and $0^\circ$ latitude, and the direction of $\bm{v}_a$ is $270^\circ$ longitude and $0^\circ$ latitude, with $|\bm{v}_a|$ comparable to the local Galactic virial velocity $\sim 10^{-3} c$ \cite{Wu2019PRL}.
By transforming to the celestial coordinate system in Fig. \ref{fig1}(b), $\bm{v}_a$ can be expressed as \cite{Wu2019PRL,Smorra2019Nature,Jiang2021NatPhys}
\begin{equation}
\bm{v}_{a} = |\bm{v}_{a}| [\cos \delta \cos\eta \hat X + \cos \delta \sin\eta \hat Y + \sin\delta \hat Z],
\end{equation}
where $\delta \approx -48^\circ$ and $\eta \approx 138^\circ$ are declination and right ascension, respectively \cite{LAMBDA}.
Therefore, the effective magnetic field \eqref{beff} has the components,
\begin{eqnarray}
B_{a,X} = \frac{2g_{\mathrm{aNN}}|\bm{v}_{a}|}{\gamma} \sqrt{2\hbar c\rho_a} \cos \delta \cos\eta \sin(2\pi\nu_a t +\phi),\\
B_{a,Y} = \frac{2g_{\mathrm{aNN}}|\bm{v}_{a}|}{\gamma} \sqrt{2\hbar c\rho_a} \cos \delta \sin\eta \sin(2\pi\nu_a t +\phi),
\end{eqnarray}
yielding the equatorial component:
\begin{equation}
B_{a,\perp} = \frac{2g_{\mathrm{aNN}}|\bm{v}_{a}|}{\gamma} \sqrt{2\hbar c\rho_a} \cos \delta \sin(2\pi\nu_a t +\phi).
\label{Baperp}
\end{equation}

Dual-species co-magnetometers provide an effective approach to detecting minute signals generated by effective magnetic fields induced by new interactions \cite{Bulatowicz2013PRL}. The colocalization of the two constituent species within the same spatial domain enables the cancellation of the background magnetic field, facilitating the detection of signals induced by new physics. This configuration allows for the extraction of sidereal modulated signals from the noisy background in precision measurements. The ultrahigh sensitivity of magnetometers to magnetic field variations offers substantial potential for exploring new physics, such as spin-gravity interactions \cite{Venema1992PRL}, nuclear EDMs \cite{Sachdeva2019PRL}, and $CPT$ and Lorentz symmetry violations \cite{Brown2010PRL}.

A $^{129}\mathrm{Xe} + ^{3}\mathrm{He}$ co-magnetometer has been applied to search for the constant cosmic background field arising from Lorentz violation \cite{Allmendinger2014PRL}.
By extracting the frequency modulation effect of Earth rotation on nuclear precession in the co-magnetometer, an upper limit on the equatorial component of the effective field
has been obtained as
\begin{equation}
B_\perp < 8.4\times 10^{-34}~\mathrm{GeV} \label{cons}
\end{equation}
at the $68\%$ confidential level (CL) \cite{Allmendinger2014PRL}.
This result can be employed to set limits on axion dark matter by interpreting it as the source of the field. Since the constraint \eqref{cons} is derived using the sidereal angular frequency $\Omega \approx 2\pi \times 1.16\times 10^{-5}$ s$^{-1}$ as a modulation, the limits are applicable only when $2\pi\nu_a$ is significantly less than $\Omega$, for instance, by an order of magnitude. The obtained upper limit applies to axion masses analyzed in this context, $m_a \leq 5\times 10^{-21}$ eV.

\begin{figure}[htbp]
    \centering
    \includegraphics[width=0.45\textwidth]{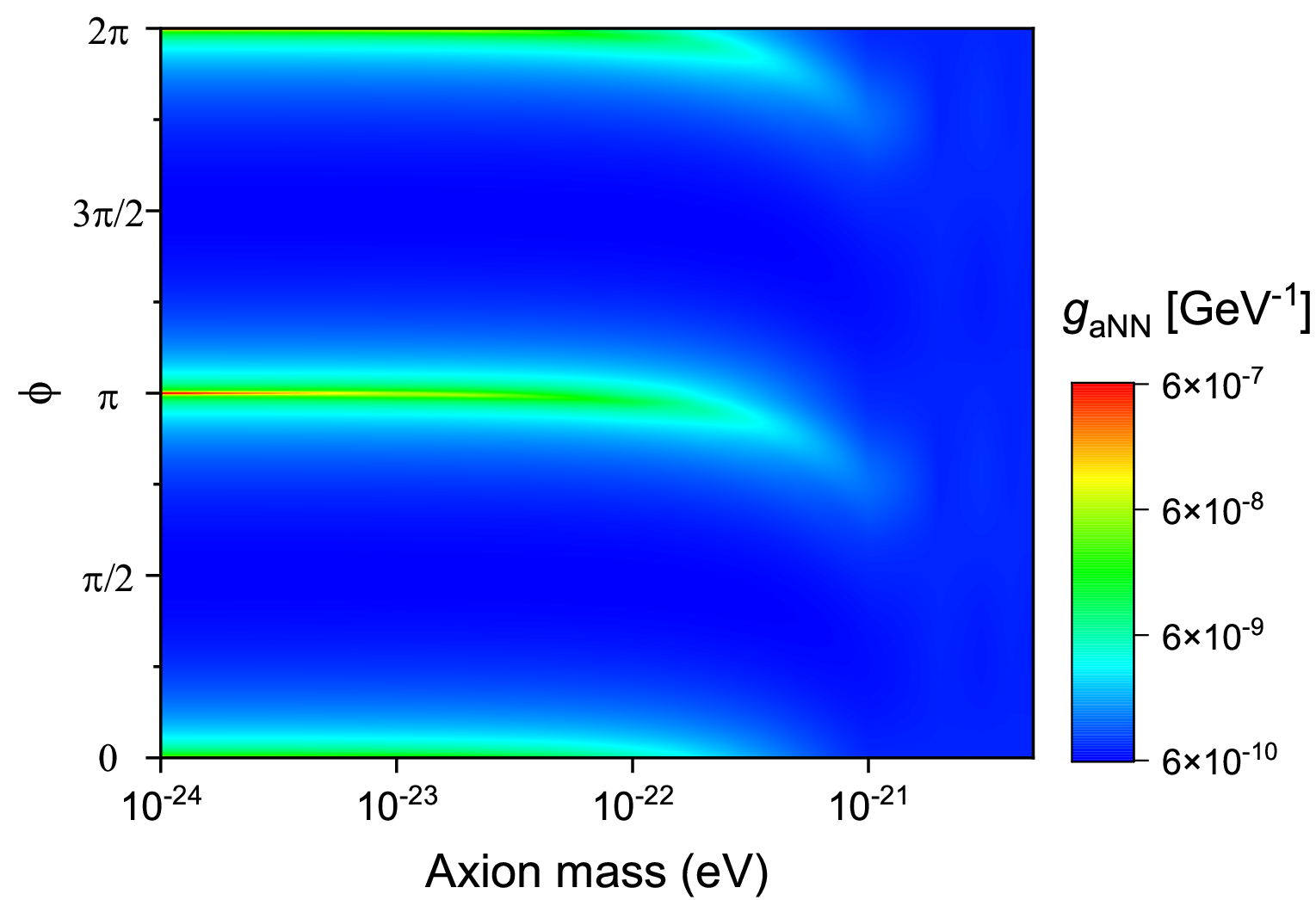}
    \caption{Likelihood analysis results for the coupling strength $g_{\mathrm{aNN}}$ of nucleons with the axion dark matter for given phase $\phi$ from 0 to $2\pi$ in the mass range $10^{-24} \lesssim m_a \lesssim 5\times 10^{-21}$ eV.}
    \label{fig2}
\end{figure}

\textit{Derived constraints.}---The experiment was conducted for approximately $10^6$ s with a sampling frequency of $1/3.2~\mathrm{s}$ \cite{Allmendinger2014PRL}, obtaining a large dataset ($N \approx 3 \times 10^5$) for the equatorial component of the effective field \cite{Allmendinger2014PRL}. Assuming white Gaussian noise with variance $\sigma^2$, a likelihood analysis for given values of $m_a$ and $\phi$ can be carried out using the following likelihood function:
\begin{equation}
\mathcal{L}(g)=\prod_{i=1}^N\frac{1}{\sqrt{2\pi\sigma^2}}\exp\left(-\frac{(b_i-g|\sin(2\pi\nu_a t_i+\phi)|)^2}{2\sigma^2}\right),
\end{equation}
where $g= 2g_{\mathrm{aNN}}|\bm{v}_{a}| \sqrt{2\hbar c\rho_a} \cos \delta/\gamma$, $b_i$ are random samples from the generalized Rice distribution that the equatorial component follows \cite{SUPP}, and $t_i$ are time series spanning from $0$ to $10^6$ s.
Maximizing the likelihood function yields the estimate for $g$,
\begin{equation}
\hat{g}=\frac{\sum_{i=1}^Nb_i|\sin(2\pi\nu_a t_i+\phi)|}{\sum_{i=1}^N\sin^2(2\pi\nu_a t_i+\phi)}.
\end{equation}
Given other known quantities such as $\cos\delta$ and $\gamma$, the obtained results for $g_{\mathrm{aNN}}$ are illustrated in Fig. \ref{fig2}.
For extreme ultralight axions with masses in the range $10^{-24} \leq m_a \lesssim 5\times10^{-23}$ eV, the function $\sin(2\pi\nu_a t +\phi)$ exhibits marginal variation over the $10^6$ s duration of the experiment.
Under these conditions, axion dark matter behaves effectively as a DC effect, and the estimate of $g_{\mathrm{aNN}}$ becomes highly sensitive to the phase $\phi$, with particularly large values occurring when $\phi$ is close to $0$, $\pi$, or $2\pi$.
For $m_a \gtrsim 5 \times 10^{-22}$ eV, the experimental duration is sufficient to capture nearly a full oscillation--or more--of the sine function, thereby reducing the dependence on the phase $\phi$.

\begin{figure}[htbp]
    \centering
    \includegraphics[width=0.45\textwidth]{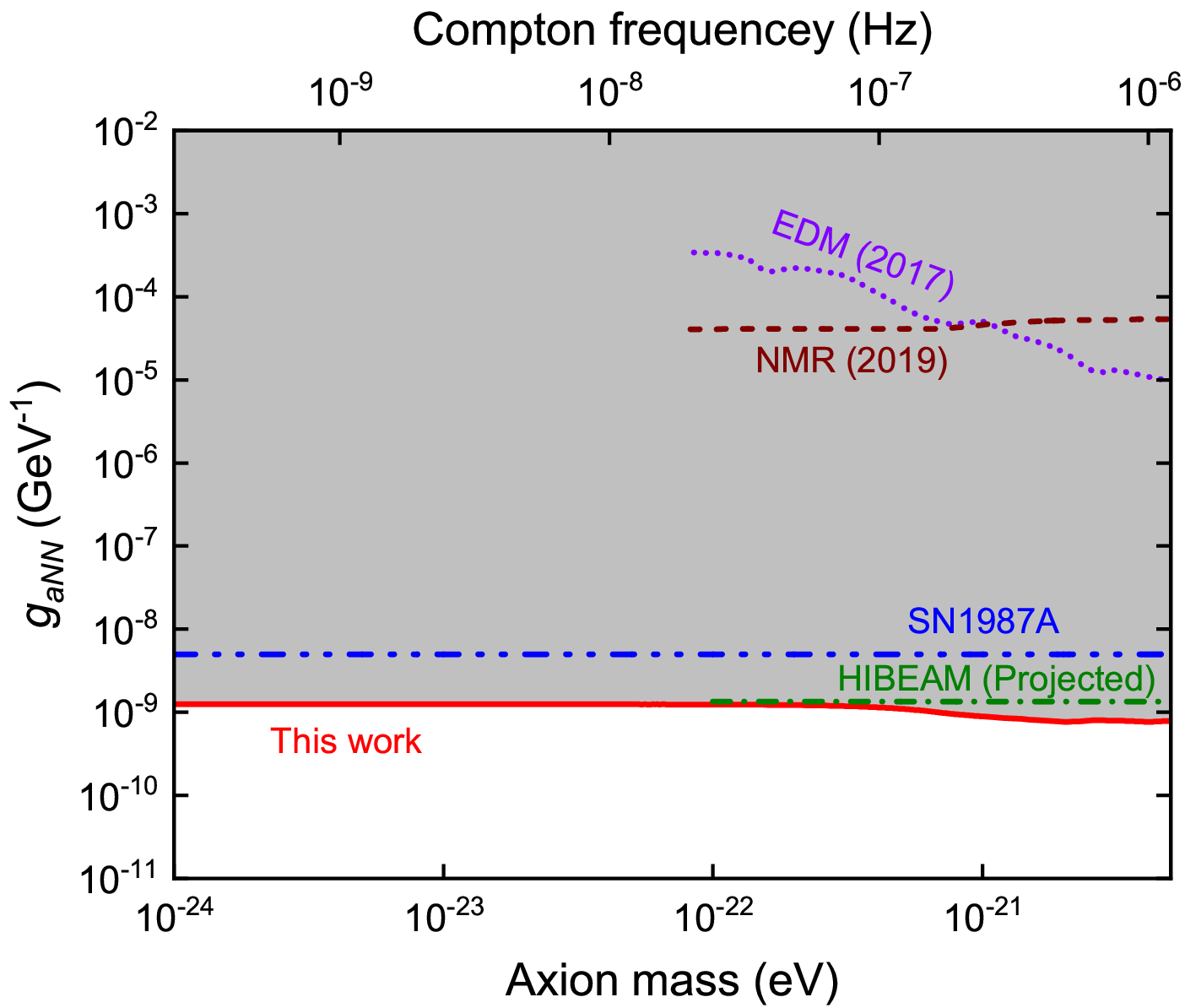}
    \caption{Limits on coupling strength $g_{\mathrm{aNN}}$ of nucleons with the axion dark matter in the mass range $10^{-24} \lesssim m_a \lesssim 5\times 10^{-21}$ eV. The red solid line represents the limits derived from the equatorial component of the effective field at the $68\%$ CL in this work, with the dark region being the parameter space excluded. The violet dotted, wine dashed, and blue dash-dot-dotted lines show the limits from the PSI neutron EDM experiment in 2017 \cite{Abel2017PRX}, the NMR-based experiment in 2019 \cite{Wu2019PRL}, and supernova SN1987A cooling \cite{Raffelt1990PhysRep, Raffelt2008}, respectively. The limits projected for a recently proposed experiment using the HIBEAM neutron beamline at the ESS \cite{Fierlinger2024PRL} are indicated by the olive dash-dotted line.}
    \label{fig3}
\end{figure}

To examine the limits on the coupling strength $g_{\mathrm{aNN}}$ in a unified manner, the uncertainty in the phase $\phi$ should be taken into account for each value of $m_a$. In the absence of prior knowledge about the phase distribution, it is reasonable to assume that $\phi$ is uniformly distributed over the interval $[0, 2\pi]$, implying that $g_{\mathrm{aNN}}$ is equally weighted across different phases. Consequently, for each $m_a$, the distribution of $g_{\mathrm{aNN}}$ as a function of $\phi$ is extracted, and the corresponding $68\%$-CL upper bound is presented in Fig.~\ref{fig3}.
The results demonstrate visible variation with $m_a$, indicating that axion dark matter cannot be approximated as a DC effect across the entire explored mass range.
Notably, the present analysis establishes the first laboratory limits exceeding the astrophysical constraints from supernova SN1987A cooling on the axion-nucleon coupling in this region.
For $10^{-22} \leq m_a \leq 5\times 10^{-21}$ eV, the limits surpass those from the PSI neutron EDM experiment (2017) \cite{Abel2017PRX} and the NMR-based experiment (2019) \cite{Wu2019PRL} by over 3 orders of magnitude.
Furthermore, the constraints are comparable to the sensitivity projected for the HIBEAM neutron beamline at the ESS in a recently proposed experiment, assuming one year of runtime \cite{Fierlinger2024PRL}.
Although the derived limits in the frequency range approximately one order of magnitude below the sidereal frequency approach those projected for the proposed experiment, the HIBEAM scheme--by combining the high precision of the Ramsey interference technique with the neutron beam's natural isolation from environmental disturbances--offers clear advantages for setting new limits in the axion mass range $5 \times 10^{-21} < m_a \le 10^{-16}$ eV. In this way, it effectively complements and extends the scope of the present work.

\begin{figure}[htbp]
    \centering
    \includegraphics[width=0.45\textwidth]{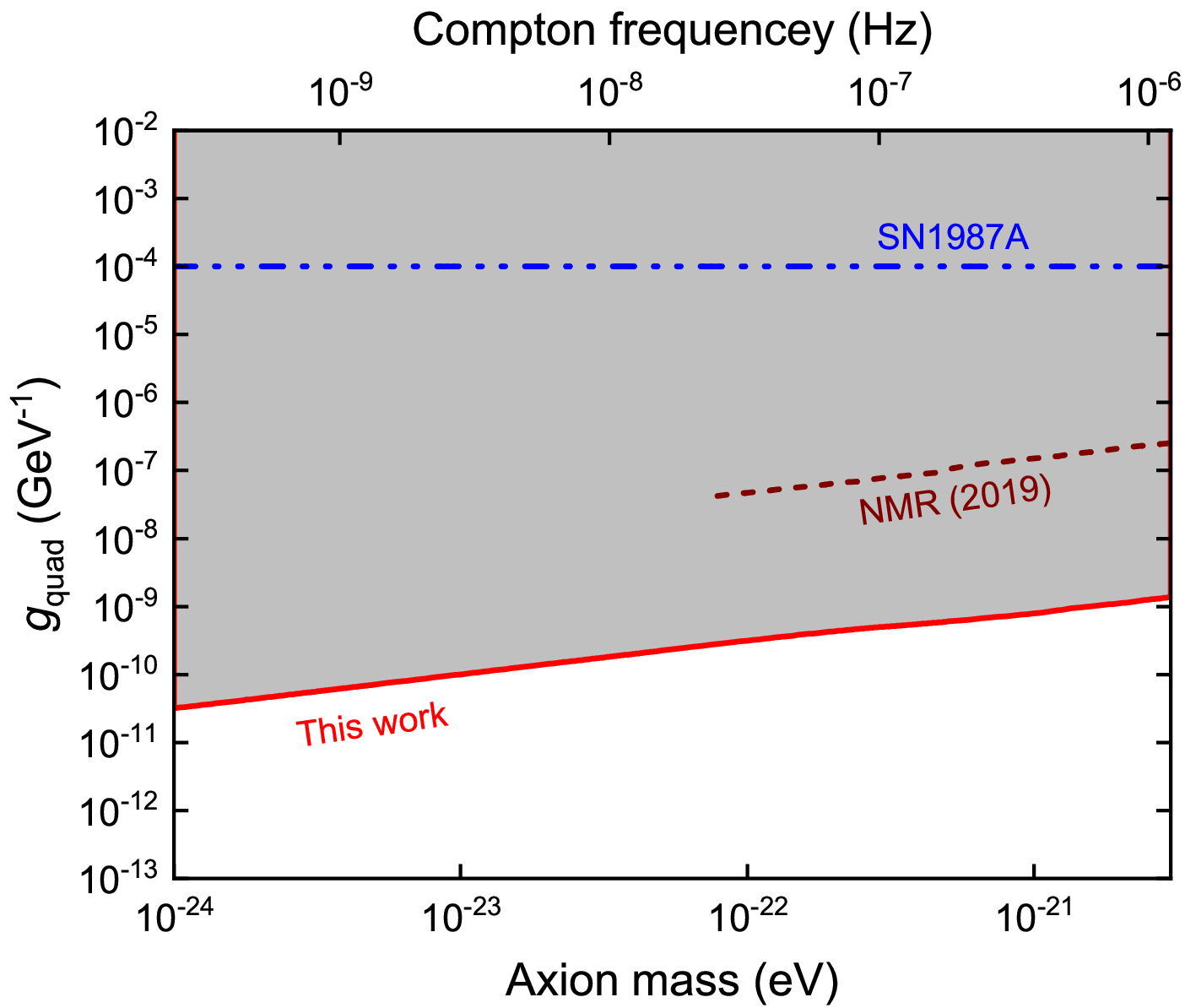}
    \caption{Same as Fig. \ref{fig3} but for axion quadratic coupling with nucleon $g_{\mathrm{quad}}$.}
    \label{fig4}
\end{figure}

Theoretically, there are scenarios in which the interaction between nuclear spins and the gradient of the axion field, $\partial_\mu a$, is suppressed, while the coupling to the gradient of the squared axion field, $\partial_\mu a^2$--known as the quadratic wind coupling--becomes dominant \cite{Olive2008PRD,Pospelov2013PRL}:
\begin{equation}
H_{\mathrm{quad}} = 2g_{\mathrm{quad}}^2 \hbar^2 c^2 \frac{\rho_a}{2\pi\nu_a}\sin(4\pi\nu_a t +\phi) \bm{v}_a \cdot \bm{I}_N,
\end{equation}
where $g_{\mathrm{quad}}$ denotes the quadratic coupling strength between axions and nuclear spins. Limits for this coupling have been derived based on similar analyses. The corresponding results are shown in Fig. \ref{fig4}. The derived limits on $g_{\mathrm{quad}}$ exceed the constraints from supernova SN1987A cooling by more than 4 orders of magnitude and improve upon the previous best laboratory limits, established by the 2019 experiment using NMR techniques, by approximately 2 orders of magnitude.

\begin{figure}[htbp]
    \centering
    \includegraphics[width=0.45\textwidth]{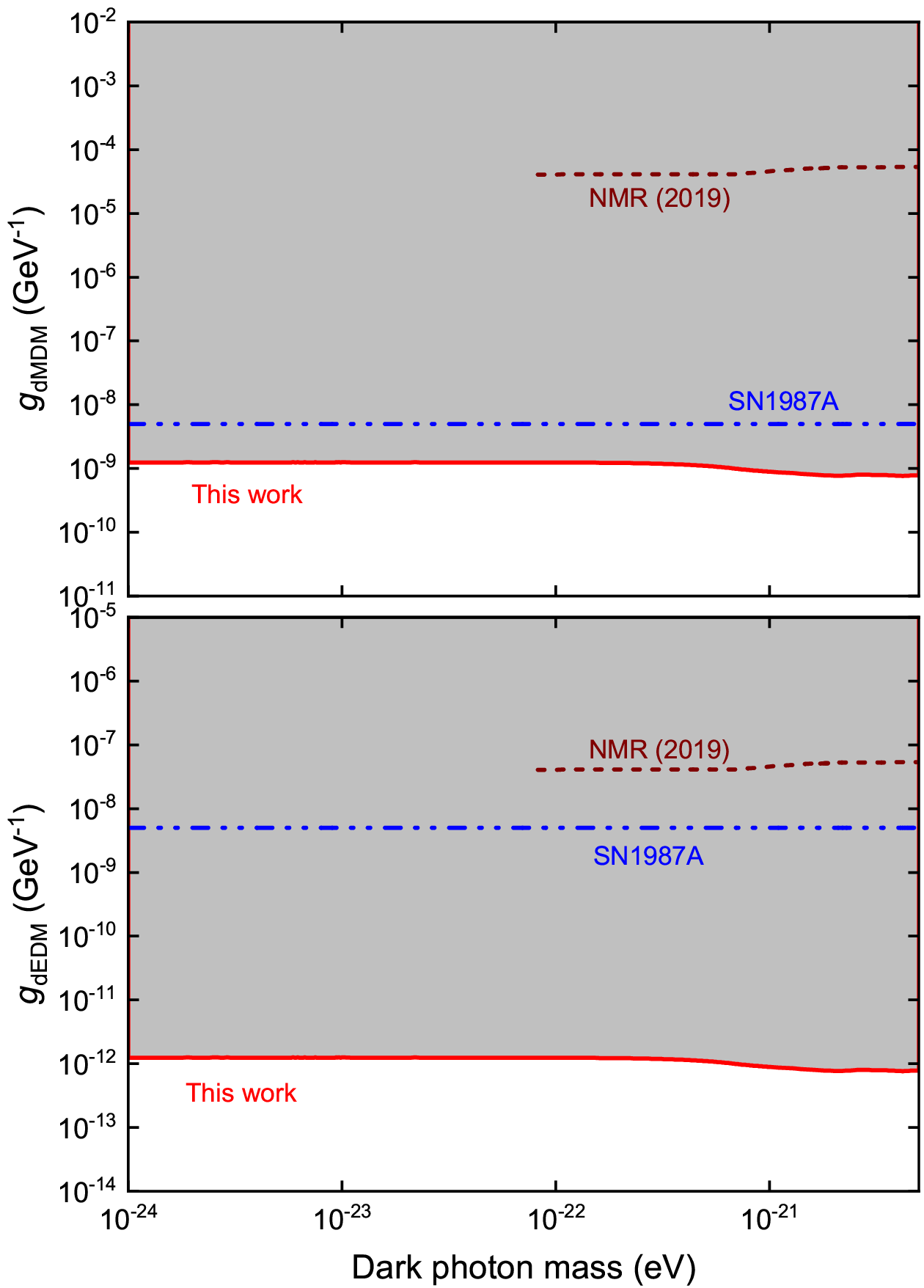}
    \caption{Same as Fig. \ref{fig3} but for dark photon-nucleon dMDM coupling (upper) and dEMD coupling (lower).}
    \label{fig5}
\end{figure}

Other possible couplings of bosonic dark matter fields, such as those mediated by dark photons--whose existence has long been conjectured and remains a topic of interest \cite{DP2021Book}--may also arise. Given the similarity to Eq. \eqref{Hint}, one can interpret the constrained effective field as originating from nuclear spin interactions with dark photons,
\begin{eqnarray}
H_{\mathrm{dMDM}} = 2g_{\mathrm{dMDM}} \sqrt{2\hbar^3 c\rho_a} \sin(2\pi\nu_a+\phi) \bm{v}_a \cdot \bm{I}_N, \label{Hm}\\
H_{\mathrm{dEDM}} = 2g_{\mathrm{dEDM}} \sqrt{2\hbar^3 c\rho_a} \sin(2\pi\nu_a+\phi) \frac{\bm{v}_a}{|\bm{v}_a|} \cdot \bm{I}_N, \label{He}
\end{eqnarray}
where $g_{\mathrm{dMDM}}$ and $g_{\mathrm{dEDM}}$ parameterize the coupling strengths between nuclear spins with the dark magnetic fields and electric fields, respectively.
Following a similar methodology to derive constraints on the axion-nucleon coupling, the resulting limits for couplings with dark photons are presented in Fig. \ref{fig5}.
Since $H_{\mathrm{dMDM}}$ \eqref{Hm} shares the same functional form as $H_{\mathrm{int}}$ \eqref{Hint}, the constraints for $g_{\mathrm{dMDM}}$ are identical to those for $g_{\mathrm{aNN}}$.
Additionally, for a given mass, the constraint for $g_{\mathrm{dEDM}}$ differs from that for $g_{\mathrm{dMDM}}$ by a factor of $10^{-3}$.
These features have also been reported in existing literature \cite{Wu2019PRL, Jiang2021NatPhys}.
As such, we refrain from further elaboration and comparison of the content in Fig. \ref{fig5}.

\textit{Conclusion and discussion.}---In conclusion, new limits on the axion-nucleon coupling have been derived over the axion mass range of $10^{-24} \leq m_a \leq 5\times 10^{-21}$ eV by reanalyzing data from laboratory measurements of Lorentz and $CPT$ violation.
These results establish the first laboratory constraints on the axion-nucleon coupling for $m_a < 10^{-22}$ eV and provide the first laboratory limits surpassing the astrophysical bounds from supernova SN1987A cooling for $10^{-22} \leq m_a \leq 5 \times 10^{-21}$ eV.

The new limits improve sensitivity by more than 3 orders of magnitude over the best existing laboratory constraints and exceed the projected reach of a recently proposed experiment utilizing high-intensity neutron beams at the ESS \cite{Fierlinger2024PRL}.
Therefore, the new results presented in this work constitute a significant advancement, especially for axion field frequencies below $10^{-8}$ Hz, where tracking the initial phase of the axion field becomes crucial, as it plays an important role in the detection of such low-frequency signals.
It appears practically impossible to effectively estimate the initial phase without first observing the axion-wind interaction in reality. Under such circumstances, the most reasonable approach is to statistically incorporate contributions from various phases and marginalize over them, as implemented in this work. In certain extreme cases--such as when the local phase is close to $0$ or $\pi$--the resulting limits for $m_a < 10^{-21}$ eV may become overly stringent.
It is also worth noting that the projected axion mass range targeted by the ESS experiment, $10^{-22}$ eV to $10^{-16}$ eV, overlaps with the mass range considered in this work only within a narrow window. In the present analysis, frequencies near or above the sidereal frequency are beyond reach, as demodulation becomes ineffective in that regime.
The future implementation of the HIBEAM experiment would therefore be highly valuable for establishing new limits on the axion-nucleon coupling in the mass range $10^{-20} \lesssim m_a \lesssim 10^{-16}$ eV.

Finally, the analysis in this work yields constraints exceeding previous results by approximately 2 orders of magnitude for the potential quadratic wind coupling.
By using similar methods, new constraints on nuclear spin interactions with dark photons have also been derived.

\textit{Acknowledgment.}---Helpful discussions with Y. M. Ma are greatly appreciated. This work was supported by the National Natural Science Foundation of China (Grants No. U2230207 and No. 12305125), the National Key Laboratory of Neutron Science and Technology (Grant No. NST202401016), and the Sichuan Science and Technology Program (Grant No. 2024NSFSC1356).

%\bibliography{Reference}
%apsrev4-2.bst 2019-01-14 (MD) hand-edited version of apsrev4-1.bst
%Control: key (0)
%Control: author (8) initials jnrlst
%Control: editor formatted (1) identically to author
%Control: production of article title (0) allowed
%Control: page (0) single
%Control: year (1) truncated
%Control: production of eprint (0) enabled
%

\end{document}